\documentclass[12pt]{spieman}  % 12pt font required by SPIE;
\usepackage{amsmath,amsfonts,amssymb}
\usepackage{graphicx}
\usepackage{setspace}
\usepackage{tocloft}

\usepackage{xcolor}
\usepackage{lineno, soul,upgreek}

\graphicspath{ {./figures/} }

\title{Bridging classical and quantum approaches in optical polarimetry: Predicting polarization-entangled photon behavior in scattering environments}

\author[a,*]{Vira R. Besaga}
\author[b,*]{Ivan V. Lopushenko}
\author[b]{Oleksii Sieryi}
\author[b]{Alexander Bykov}
\author[a,c]{Frank Setzpfandt}
\author[d]{Igor Meglinski}
\affil[a]{Friedrich Schiller University Jena, Institute of Applied Physics \& Abbe Center of Photonics, Albert-Einstein-Str. 6, Jena, 07745, Germany}
\affil[b]{University of Oulu, Opto-Electronics and Measurement Techniques, P.O. Box 4500, Oulu, FI-90014, Finland}
\affil[c]{Fraunhofer Institute for Applied Optics and Precision Engineering IOF, Albert-Einstein-Str. 7, Jena, 07745, Germany}
\affil[d]{Aston University, College of Engineering and Physical Sciences, Birmingham, B4 7ET, UK}

\cftpagenumbersoff{figure}
\cftpagenumbersoff{table} 
\begin{document} 
\maketitle

\begin{abstract} We explore quantum-based optical polarimetry as a potential diagnostic tool for biological tissues by developing a theoretical and experimental framework to understand polarization-entangled photon behavior in scattering media. We investigate the mathematical relationship between Wolf's coherency matrix in classical optics and the density matrix formalism of quantum mechanics which allows for the extension of classical Monte Carlo method to quantum states. The developed generalized Monte Carlo approach uniquely integrates the Bethe-Salpeter equation for classical scattering, the Jones vector formalism for polarization, and the density matrix approach for quantum state representation. Therefore, this unified framework can model both classical and quantum polarization states, handle multi-photon states, and account for varying degrees of entanglement. Additionally, it facilitates the prediction of quantum state evolution in scattering media based on classical optical principles. The validity of the computational model is experimentally confirmed through high-fidelity agreement between predicted and measured quantum state evolution in tissue-mimicking phantoms. This work bridges the gap between classical and quantum optical polarimetry by developing and validating a comprehensive theoretical framework that unifies these traditionally distinct domains, paving the way for future quantum-enhanced diagnostics of tissues and other turbid environments.
\end{abstract}

% Include a list of up to six keywords after the abstract

% Include email contact information for corresponding authors
{\noindent \footnotesize\textbf{*}Correspondence to \linkable{vira.besaga@uni-jena.de} and \linkable{ivan.lopushenko@oulu.fi}. These two authors contributed equally to this work.
}

%\begin{spacing}{2}   % use double spacing for rest of manuscript

\section{Introduction} \label{sec:intro}

Whether atmospheric aerosols, particulate matter, or biological tissues, each of the mentioned examples can be described as a turbid scattering medium. Optical metrology of such substances proved substantial for maintaining free-space optical communication links~\cite{Ren:2013,Nikulin:2024}, routine environmental monitoring~\cite{Davis2005}, or characterization of biomedical samples~\cite{MCREv2017}, and, at the same time is challenged by the high optical losses of the probing light. This happens mainly due to a wide range of angles within which the photon can change its propagation direction upon each scattering event. Under such conditions, interpretation of the transmitted scalar intensity modulation induced by the sample often resigns in favor of vectorial polarization-based sensing, which implies examining the polarization response of an object/medium of interest and is known as optical polarimetry~\cite{Azzam:2016}. Among others, this technique offers insights into such internal characteristics of the specimen as chirality, anisotropy, and morphology~\cite{He:2021,Borovkova2022,Borovkova2020a}, which are valuable for technical inspection, biomedical diagnostics, remote sensing and other fundamental and applied problems.

Nowadays, intensive research activities are undertaken to boost the performance of optical metrology via employment of non-classical states of light. Examples of leveraging quantum light include the possibility to surpass the shot noise in imaging using two-photon correlated states \cite{Samantaray:2017}, increasing the phase measurement sensitivity with squeezed states \cite{Israel:2014}, and imaging with photons never interacting with the sample \cite{Lemos:2014}. Recent years have shown a growing interest in application of quantum light also for polarization-based sensing. Here, one should mention the development of a general theoretical framework \cite{Goldberg:2022}, experimental realization of nonlocal polarimetry \cite{Restuccia:2022,Magnitskiy:2022}, and the evaluation of sensitivity enhancement in polarization sensing with NOON states \cite{Wolfgramm:2013,Belsley:2022,Pedram:2023}. Particularly attractive is the employment of polarization-entangled photons, which find extensive interest in quantum communication and sensing~\cite{Zhang:2024:SciAdv,Liu:2024}. Among findings relevant for biomedical diagnostics, it is worth mentioning theoretical prediction of entanglement conservation through a scattering medium \cite{Velsen:2004} and first experiments indicating deeper penetration into a biological tissue \cite{Shi:2016}. In our earlier works, we have proved the applicability of polarization-entangled photons for sensing of samples with subtle polarization response including monolayer cell cultures and aqueous solutions of microorganisms \cite{Besaga:2024SPIE,Zhang:2024} and demonstrated enhanced precision of polarimetric measurement enabled by utilization of the entangled photon pairs \cite{Pedram:2024}. Recently, we have also reported on a framework towards nonlocal polarization-based classification with substantially reduced number of measurements per sample \cite{Vega:2021, Besaga:2024APLphotonics}. 

Nevertheless, quantum polarimetry still remains in the dawn of its establishment. In pursuit of real-life applications of quantum enhanced polarimetry for the quantitative diagnostics of practically relevant samples like biomedical tissues or other turbid environments, the fundamental understanding of the mechanisms behind the evolution of quantum states of light upon interaction with the sample under study is highly demanded and is not yet available. Taking into account the broad employment of polarization-entangled photons, the studies of such photonic states stand out as particularly important. Here, the possibility to interpret and predict the behavior of the probing photonic state dependent on the sample's optical properties is a prerequisite for establishment of optimal measurement algorithms and definition of appropriate metrological metrics.

In the present study, we focus on the field of biomatter-light interaction and refer to Monte Carlo (MC) modelling to approach the problem of fundamental understanding, interpretation and prediction of the polarization-entangled photons behavior driven by the medium under study. MC techniques are widely recognized as efficient tools for studying light transfer within biological tissues and other examples of turbid medium~\cite{Lopushenko2024,MCissue_JBO2022,Xu2004,Ghosh:2011} that find applications in such hot topics as early cancer diagnostics~\cite{Kuchimaru2022}, photodynamic therapy~\cite{Wang2022}, and many others~\cite{MCREv2017,Meglinski2024}.
Particularly, we refer to the Bethe-Salpeter equation (BSE) to track the polarization state of light attenuated by the turbid medium~\cite{RoyalSoc2005,Kuzmin2007,TuchinBookMeglinski2013,Lopushenko2024}. This method, allowing to overcome the problem of polarized light tracing impossible in the early MC models~\cite{Wang1995}, co-exists with the treatment of the polarized light via the vector radiative transfer equation (VRTE)~\cite{Ramella2005pt1,Wang2001,Xu2004,Muinonen2004}. Both VRTE and BSE approaches ensure independent comprehensive means to follow photon polarization evolution induced by scattering, while exhibiting a fundamental relation between each other which has only recently been established~\cite{DoicuMishchenko2019}. The BSE framework appears especially notable for modelling polarization-entangled photonic states, since it inherently allows to trace different polarization states of each photon along its trajectory, simultaneously and independently. Another advantage of the BSE-based MC modelling is its close connection to the Jones vector formalism~\cite{Akkermans1988,Kuzmin,Lopushenko2024}. It implies the possibility to directly evaluate Wolf's coherency matrix both for single photons and for photon ensembles and allows intuitive physical interpretation of the multiple scattering process via ladder diagrams~\cite{RoyalSoc2005,Kuzmin}. 

MC techniques have been already extensively used for addressing nontrivial quantum problems. The family of approaches known as Quantum Monte Carlo (QMC) methods have been employed to estimate the quantum state of many-body systems and investigate the behavior of complex quantum ensembles via wave function sampling \cite{Kim2018,Carlson2015}. 

In this study, in contrast to QMC, we employ the MC principles to mimic the scattering process in a quantum channel and introduce a new model to expand the applicability of the approach to prediction of any two-photon quantum state evolution after interaction with a turbid medium. As an example, we consider a Bell state in the form $\vert\Psi^+\rangle = 1/\sqrt{2}(\vert HV \rangle+\vert VH \rangle$) and scattering of one of its partner photons within a tissue mimicking phantom. The latter imitates the behavior of a real biomedical sample~\cite{Sieryi:2020} whereas the exploited scenario can be translated to atmospheric aerosols, particulate matter, and other scattering media. We discuss in detail the performance of MC modelling for the described scenario and prediction of the photonic state evolution. We support our theoretical findings with a corresponding experimental demonstration. Hereby, the choice of the photonic state and sample within the study and the presented results appear interdisciplinary and bridging the fields of turbid medium optics and quantum enhanced technologies. This emphasizes the impact of the presented study and its importance for perspective applications ranging from biomedical metrology to quantum communication and remote sensing.

\section{Model} \label{sec:Methods}

Within a turbid scattering medium, each photon can follow a plethora of different trajectories defined by the material properties. We simulate these with the MC algorithm which combines aspects of the Bethe-Salpeter equation~\cite{DoicuMishchenko2019} and radiative transfer theory~\cite{Mishchenko2002}. The algorithm and its physical background has been extensively covered in our previous works~\cite{Lopushenko2024,TuchinBookMeglinski2013,Kuzmin,RoyalSoc2005} and is also described in Section S1 of the Supplementary Material. Importantly, the photons within MC modelling should not be misinterpreted as physical photons, but are rather considered as statistical particles obeying the radiative transfer equation, which has the formal mathematical structure of a kinetic equation 
describing particle transport. For this reason, in the current paper, the photons within MC model will be also referred to as photon packets, and physical photons will be referred to as photons.

\begin{figure}[!t]
\centering
\includegraphics[width=0.85\linewidth]{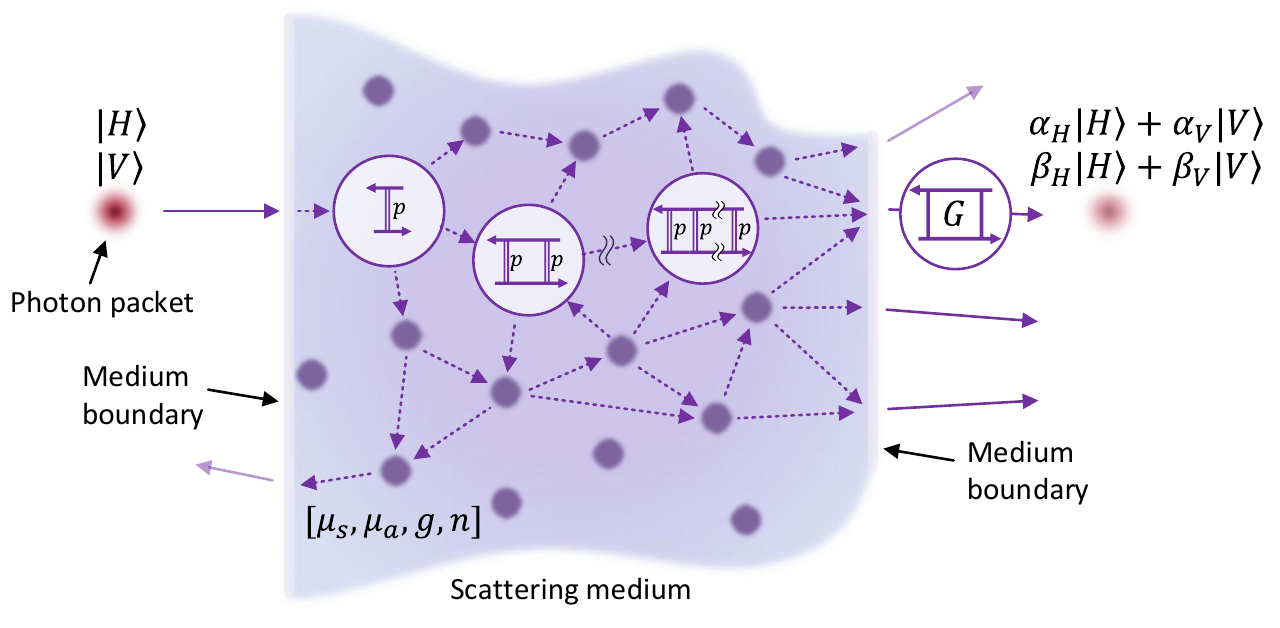}
\caption{Schematic representation of possible trajectories of the photon packets passing through a scattering medium. Notation $\vert H \rangle$ and $\vert V \rangle$ addresses the problem of entangled state modelling, where each trajectory can be followed by a photon with both horizontal $\vert H\rangle$ and vertical $\vert V\rangle$ polarization. In turn, $\alpha_{H}\vert H\rangle + \alpha_{V}\vert V\rangle$ and $\beta_{H}\vert H\rangle + \beta_{V}\vert V\rangle$ correspond to the resultant polarization states for initial either $\vert H \rangle$ or $\vert V \rangle$ input states (for details on such notation refer to Sec.~\ref{sec:polTracing} and Sec.~\ref{sec:simApproaches}). $\mu_s$, $\mu_a$, $g$, and $n$ define the properties of the medium: scattering coefficient, absorption coefficient, scattering anisotropy factor, and refractive index, correspondingly~\cite{Tuchin2015book}. One of the trajectories corresponds to a snake photon path and features ladder diagrams for visualization of the iterative solution of the Bethe-Salpeter equation. Here, $G$ denotes the propagator of the Bethe-Salpeter equation and $p$ stands for the scattering phase function (adapted from Ref.~\cite{RoyalSoc2005}).}
\label{fig:MCmodelling}
\end{figure}

The key steps of our model include: 1) the launch of a large amount ($N_{inc}>10^{9}$) of photon packets from the light source; 2) interaction with the sample with predefined scattering coefficient $\mu_s$, absorption coefficient $\mu_a$, scattering anisotropy factor $g$, and refractive index $n$~\cite{Tuchin2015book} (see Fig.~\ref{fig:MCmodelling} for artistic illustration of different possible scattered trajectories); and 3) collecting the statistics from the $N_{ph}<N_{inc}$ photon packets which arrived at the detector~\cite{Lopushenko2024}. Each photon packet is supplied with a statistical weight $W_j$, $j=[1 ... N_{ph}]$, which is proportional to its intensity, and with the polarization state. In this work, a uniform intensity distribution is simulated, which leads to unit initial weight for each photon packet. In the course of propagation through the turbid medium, the statistical weight of each packet is attenuated with respect to the Beer-Lambert law along its trajectory. After launch, a photon packet begins propagation according to the defined initial direction $\mathbf{s}$, which is updated after each scattering event with respect to the scattering phase function, or after an interface collision event with respect to Snell law. Photon packets satisfying the detection conditions are contributing to the final statistics.

Next, we will address in detail polarization state tracing along the trajectory of each launched photon packet by using the Jones vector framework. Consequently, we will utilize Wolf's coherency matrix to express the polarization state in our MC model in the matrix form. Finally, we will refer to the fundamental relation between coherency and density matrices to further generalize the MC model in a way which allows consideration of polarization-entangled photon pairs.

\subsection{Polarization state tracing}
\label{sec:polTracing}

The polarization tracing mechanism which is invoked during the interaction of the photon packets with the sample is of the utmost interest. In this study, to describe the polarization state we employ the Jones vector $\boldsymbol\varepsilon = (E_x, E_y)^T$ in the notation adopted from Mandel and Wolf~\cite{Wolf1995} for the polarization vector. In the present work this vector is not necessarily unit and fully defines the electric field vector $\mathbf{E}=(E_x,E_y,0)^T$ allowing for the description of any polarization state of fully polarized light. Here and onwards, $T$ corresponds to matrix transpose. By fully polarized state of light we explicitly mean that strict equality holds in the following relation between elements of Stokes vector $\mathbf{S}=(S_0, S_1, S_2, S_3)^T$: $S_0^2 = S_1^2+S_2^2+S_3^2$.~\cite{Born2019}

Any polarization state can be expressed via combination of the two basis vectors. Here, we use a basis of horizontal $\boldsymbol\varepsilon_H$ and vertical $\boldsymbol\varepsilon_V$ polarization states:
\begin{equation}
\label{eq:basisDecompositionJ}
\boldsymbol{\varepsilon} = E_{x} \boldsymbol{\varepsilon}_H + E_{y} \boldsymbol{\varepsilon}_V = E_{x} \left(\begin{array}{c}
    1\\ 
    0\\ 
    \end{array}\right) + E_{y} \left(\begin{array}{c}
    0\\ 
    1\\ 
    \end{array}\right).
\end{equation}
This representation allows to assign any input polarization state to the launched photon packet. To properly track the evolution of the polarization state of photon packets that undergo scattering and are later detected, we introduce the three-component polarization vector $\mathbf{P}$ which corresponds to the $\mathbf{E}$ field direction~\cite{Akkermans1988,RoyalSoc2005,Kuzmin,TuchinBookMeglinski2013}. Correspondingly, each photon packet trajectory, which contains $N$ scattering events, at start is supplied with a $\mathbf{P}_0$ vector representing its initial polarization state. In case of the $\boldsymbol\varepsilon_H$ state, $\mathbf{P}_0=(1,0,0)^T$, and for the $\boldsymbol\varepsilon_V$ state $\mathbf{P}_0=(0,1,0)^T$. Then, within the iterative solution to the BSE~\cite{RoyalSoc2005,Kuzmin2007}, the evolution of this vector can be traced along the photon packet trajectory:~\cite{Akkermans1988,TuchinBookMeglinski2013,Lopushenko2024}
\begin{equation}
 \label{eqn:SvMC_BSprocedure_expanded}
    \mathbf{P}_{N} = \hat{\mathbf{U}}_N \hat{\mathbf{U}}_{N-1} \hat{\mathbf{U}}_{N-2} ... \hat{\mathbf{U}}_{1} \mathbf{P}_{0}, \quad \hat{\mathbf{U}}_i = -\mathbf{s}_i\times\left[\mathbf{s}_i\times\mathbf{P}_{i-1}\right].
\end{equation} 
Here, $\mathbf{P}_N$ corresponds to the polarization state of the photon packet that has arrived on the detector and $\mathbf{s}_i$ corresponds to the photon packet direction after the $i$-th scattering event. With $\mathbf{P}_N$ value obtained, the final polarization state of each photon packet can be reconstructed in the form of a Jones vector: there always exists a one-to-one correspondence between $\mathbf{P}$ and $\boldsymbol\varepsilon$, which is detailed in the Supplementary Material, S1.3. It is important to note that individual photon packets remain fully polarized after each scattering or surface interaction event~\cite{Lopushenko2024}. This fact enables application of the Jones formalism to describe polarization state of the scattered photon packets. We also note that this approach works within the Rayleigh-Gans-Debye approximation. The latter assumes that the medium turbidity is conditioned by the presence of optically soft particles in it, which means that the refractive index of each scatterer is close to that of the surrounding medium~\cite{Akkermans1988,TuchinBookMeglinski2013,Bohren1983}. 

The final goal of this study is modelling of polarization-entangled photons on the example of a Bell state in the form $\vert\Psi^+\rangle = 1/\sqrt{2}(\vert HV \rangle+\vert VH \rangle$). For this reason, we supply each photon packet at launch with either horizontal $\boldsymbol\varepsilon_H$ or vertical $\boldsymbol\varepsilon_V$ polarization. These states are equivalent to the states $\vert H \rangle = (1, 0)^T$ and $\vert V \rangle = (0, 1)^T$ expressed in the Dirac notation. We will discuss the relationship between $\boldsymbol\varepsilon$ and the corresponding ket vector $\lvert \cdot \rangle$ in the subsequent sections. By assigning a pair of independent vectors $\mathbf{P}_x, \mathbf{P}_y$ to each photon packet we are able to trace two separate polarization states along the same trajectory. This plays an especially important role within the task of modelling the entangled photons' behavior, since until the detection of one of the partner photons the polarization state of both of them is not defined. 

For each photon packet launched with $\boldsymbol\varepsilon_H$ \big($\boldsymbol\varepsilon_V$\big) polarization, we denote its final polarization state at the detector as $\mathbf{X}$ \big($\mathbf{Y}$\big), and decompose it into the basis components with the weight factors $\mathfrak{m}$ and $\mathfrak{n}$ ($\mathfrak{p}$ and $\mathfrak{q}$) according to (\ref{eq:basisDecompositionJ}): 
\begin{equation}
\label{eq:X}
\mathbf{X} = \mathfrak{m}\boldsymbol\varepsilon_H + \mathfrak{n}\boldsymbol\varepsilon_V, \quad
\mathbf{Y} = \mathfrak{p}\boldsymbol\varepsilon_H + \mathfrak{q}\boldsymbol\varepsilon_V.
\end{equation}

\noindent Hereby, our MC model allows to simulate the behavior of Jones vector expressed via coefficients $\mathfrak{m}, \mathfrak{n}$ or $\mathfrak{p},\mathfrak{q}$ in the scattering medium for any photon packet trajectory. 

\subsection{Describing polarization: from classical framework to density matrix}
\label{sec:bridging}

A major advantage of the Jones calculus is that it can be employed to describe the corresponding pure quantum polarization states~\cite{Kwiat2001,Alfano2023}: 
\begin{equation}
\label{eq:ketDefinition}
\boldsymbol\varepsilon = \left(\begin{array}{c}
    E_{x}\\ 
    E_{y}\\ 
    \end{array}\right) \rightarrow \left(\begin{array}{c}
    \psi_1^{}\\ 
    \psi_2^{}\\ 
    \end{array}\right) = \vert \psi \rangle.
\end{equation}
Here, $\vert \psi \rangle$ is a pure quantum polarization state of an individual photon. The important difference between the two descriptions resides in the fact that the wavefunction of a quantum state describes probabilities, while the Jones vector describes the electric field. By using an arrow sign we point out that there exists a surjective relation which connects electric field components to the corresponding probability amplitude values. Below we explicitly define this relation in more general terms of coherency and density matrices.

In particular, we note that when applying the definition of the density matrix of the pure state~\cite{Kwiat2001,Landau}
\begin{equation}
\hat{\rho} = \vert \psi \rangle \langle \psi \vert
\label{eq:densityMatrixDefinition}
\end{equation}
to the Jones vector $\boldsymbol\varepsilon$, one immediately arrives at the expression for Wolf's coherency matrix of fully polarized light:~\cite{Wolf1959,Born2019}
\begin{equation}
\hat{\rho}_\mathrm{Wolf} = \vert \boldsymbol\varepsilon \rangle \langle \boldsymbol\varepsilon \vert = \boldsymbol\varepsilon \big(\boldsymbol\varepsilon^T\big)^* = \left(\begin{array}{c}
    E_{x}\\ 
    E_{y}\\ 
    \end{array}\right) \left( \begin{array}{cc}E_{x}^* & E_{y}^* \end{array}\right) = \left(\begin{array}{cc}
    E_{x} E_{x}^* & E_{x} E_{y}^*\\ 
    E_{y} E_{x}^* & E_{y} E_{y}^*\\ 
    \end{array}\right).
    \label{eq:wolfmatrix}
\end{equation}
Here, ket vector $\lvert \cdot \rangle$ is defined in the Hilbert space and bra vector $\langle \cdot \vert$ is defined as an element of the conjugated Hilbert space $\langle \cdot \vert = \left(\vert \cdot \rangle^T \right)^*$, where $^*$ corresponds to complex conjugation. For our model, it means that any photon packet with known polarization state~(\ref{eq:X}) can be supplemented with the related coherency matrix. In the following, we will apply bra and ket notation for Jones vector $\boldsymbol\varepsilon$ in the meaning of Eq.~(\ref{eq:wolfmatrix}) when considering evaluation of intensity projections. 

In biophotonics, intensity projections on the selected polarizer states and the Stokes vector of the polarized light are commonly evaluated as separate parameters~\cite{Lopushenko2024,Kuzmin}. Within the framework proposed in this work, intensity projection on any polarizer state $\boldsymbol\varepsilon$ can be evaluated if the coherency matrix of the photon packet is known:
\begin{equation}
\label{eq:intensityProjection}
I_{\boldsymbol\varepsilon} = \langle \boldsymbol\varepsilon \vert \hat{\rho}_\mathrm{Wolf} \vert \boldsymbol\varepsilon \rangle.
\end{equation}
This expression coincides with the way of evaluating quantum state observable values, which in terms of measurement correspond to the number of photons $n$ counted by a detector and which is proportional to the classical intensity~\cite{Kwiat2001,Alfano2023,Wolf1995}
\begin{equation}
{n}_{\upphi} = \mathcal{N} \langle \upphi \vert \hat{\rho} \vert \upphi \rangle,
\label{eq:intensityFromDensityMatrixDefinition}
\end{equation}
where $\upphi$ is a chosen projection state described by Eq.~(\ref{eq:ketDefinition}) and $\mathcal{N}$ is a constant that depends on light intensity and detector efficiency. In the well-known work by James et al. \cite{Kwiat2001}, tomographic measurement of the density matrix of an ensemble of single photons is demonstrated to be equivalent to the measurement of several light intensity components, as well as to Stokes parameters evaluation. We rewrite the expressions (2.3) and (2.12) from that paper in terms of the sought relation between density matrix and coherency matrix:
\begin{equation}
\hat{\rho} = \hat{\rho}_\mathrm{Wolf} / \mathrm{tr}\left(\hat{\rho}_\mathrm{Wolf}\right).
\label{eq:densityVScoherencyRelation}
\end{equation}
Here, $\mathrm{tr}(\cdot)$ corresponds to the matrix trace. Similarly, light intensity measurements in Mueller matrix polarimetry can be considered as direct counterpart of quantum process tomography~\cite{Mohseni:2008}.

The established fundamental expressions~(\ref{eq:wolfmatrix}) and~(\ref{eq:densityVScoherencyRelation}) relate the pure photonic quantum state and fully polarized state of light in classical interpretation, therefore paving the way to naturally generalize our model by expressing polarization state of the MC photon packet in both coherency and density matrix terms. 
This, in turn, enables prediction of the non-classical states of light affected by the turbid medium and thus bridges the fields of quantum optics and classical polarimetry of turbid media.

\subsection{Simulation of polarization-entangled photon pairs} \label{sec:simApproaches}

The quantum bra-ket framework can be naturally expanded to a separable pair of photons~\cite{Kwiat2001}:
\begin{equation}
\vert \Psi \rangle = \left(\begin{array}{c}
    \Psi_1\\ 
    \Psi_2\\ 
    \Psi_3\\
    \Psi_4\\
    \end{array}\right) = \left(\begin{array}{c}
    \psi^{(1)}_1 \psi^{(2)}_1\\ 
    \psi^{(1)}_1 \psi^{(2)}_2\\ 
    \psi^{(1)}_2 \psi^{(2)}_1\\
    \psi^{(1)}_2 \psi^{(2)}_2\\
    \end{array}\right) = \vert \psi^{(1)} \rangle \otimes \vert \psi^{(2)} \rangle = \vert \psi^{(1)}\rangle \vert \psi^{(2)} \rangle = \vert \psi^{(1)} \psi^{(2)} \rangle.
    \label{eq:qubitState}
\end{equation}
Here, $\vert \psi^{(i)} \rangle = \left(\psi^{(i)}_1, \psi^{(i)}_2\right)^T$ corresponds to the pure state of a single photon defined according to~(\ref{eq:ketDefinition}), and~$\otimes$~is a tensor product. 

In the following, we consider polarization-entangled photon pairs. The state of such system can not be decomposed into states of separate photons like in Eq.~(\ref{eq:qubitState}). Pure two-photon polarization-entangled states in the $H$-$V$ basis are Bell states:
$$
\begin{array}{ll}
    \vert\Phi^+\rangle = \dfrac{1}{\sqrt{2}}(\vert HH\rangle+\vert VV\rangle) = \dfrac{1}{\sqrt{2}} \left(\begin{array}{c}
    1\\ 
    0\\ 
    0\\
    1\\
    \end{array}\right), &
    \vert\Phi^-\rangle = \dfrac{1}{\sqrt{2}}(\vert HH\rangle-\vert VV\rangle) = \dfrac{1}{\sqrt{2}} \left(\begin{array}{c}
    1\\ 
    0\\ 
    0\\
    -1\\
    \end{array}\right),\\ 
    \vert\Psi^+\rangle = \dfrac{1}{\sqrt{2}}(\vert HV\rangle+\vert VH\rangle) = \dfrac{1}{\sqrt{2}} \left(\begin{array}{c}
    0\\ 
    1\\ 
    1\\
    0\\
    \end{array}\right), &
    \vert\Psi^-\rangle = \dfrac{1}{\sqrt{2}}(\vert HV\rangle-\vert VH\rangle) = \dfrac{1}{\sqrt{2}} \left(\begin{array}{c}
    0\\ 
    1\\ 
    -1\\
    0\\
    \end{array}\right).\\
    \end{array}
$$
\noindent The density matrix of a mixed state of polarization-entangled photon pair can be decomposed into a combination of density matrices of Bell states, with weight factors indicating probabilities of these states~\cite{Landau}:
\begin{equation}
\label{eq:densityMatrixDecompositionIntoBell}
\hat{\rho} = {p}_1 \vert \Psi^{+} \rangle \langle \Psi^{+} \vert + {p}_2 \vert \Psi^{-} \rangle \langle \Psi^{-} \vert + {p}_3 \vert \Phi^{+} \rangle \langle \Phi^{+} \vert + {p}_4 \vert \Phi^{-} \rangle \langle \Phi^{-} \vert.
\end{equation} 
This expression will be later employed for interpretation of the experimentally measured states. In this paper, we focus on the example of the $\vert \Psi^{+} \rangle$ state and its density matrix while the developed model is in generally applicable to any two-photon polarization state.

\begin{figure}[!t]
\centering
\includegraphics[width=0.9\linewidth]{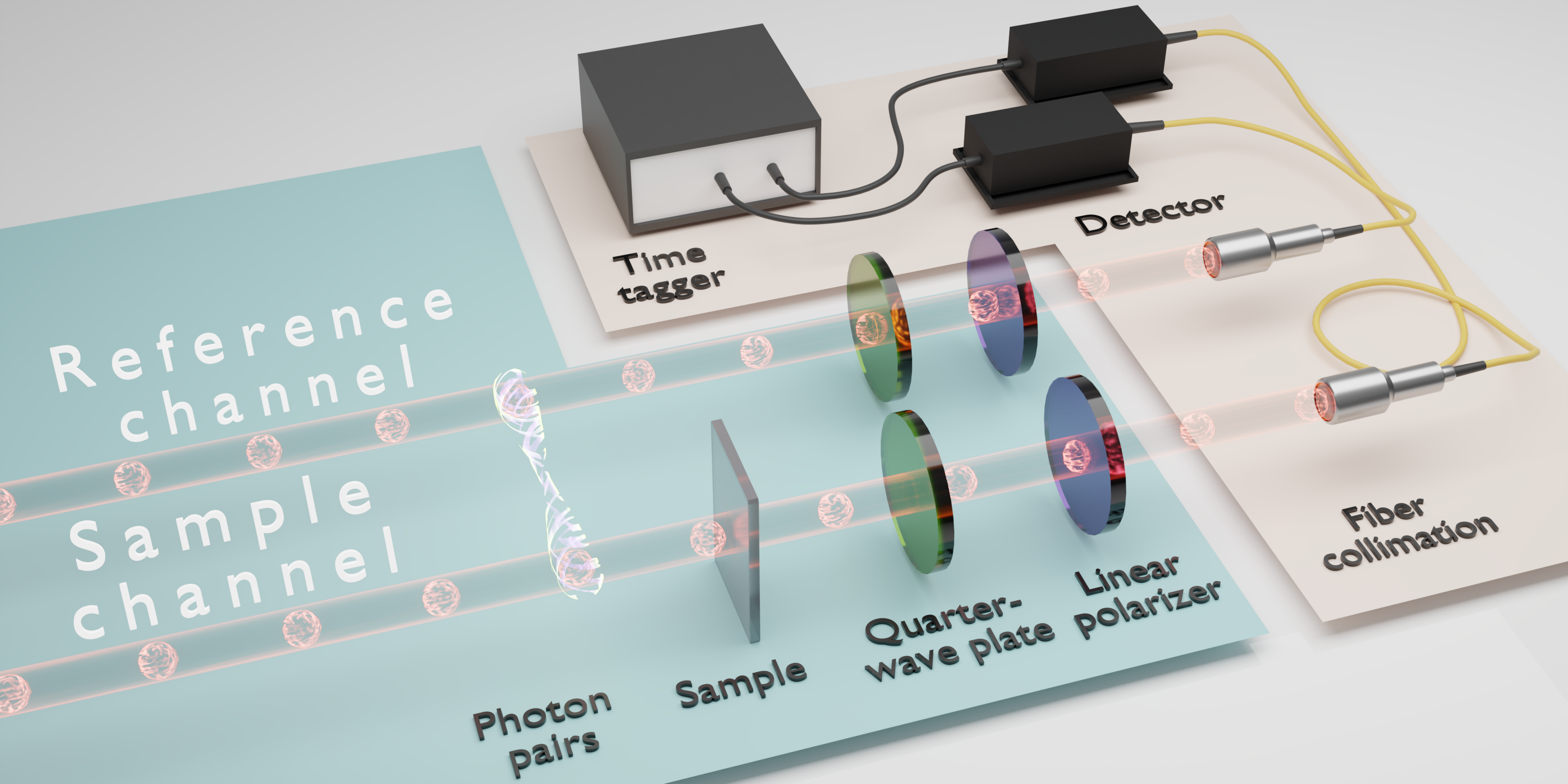}
\caption{Scattering scenario considered in this study, conceptual sketch. A pair of polarization-entangled photons is guided so that one of the partner photons interacts with a turbid medium. Another photon remains unchanged. Quarter-wave plates and linear polarizers enable polarization projective measurements for two-photon state reconstruction upon detection and coincidence events counting.}
\label{fig:optSetup} 
\end{figure}

Here, we consider the following scenario: two photons propagate in separate channels, with one of the channels containing the measured sample. The concept of such case is schematically visualized in Fig.~\ref{fig:optSetup}. The figure depicts also the implementation of the polarization projective measurements and detection scheme used in the experiments, which are described in detail in in Supplementary Material, S2.2. In terms of the state $\vert \Psi^{+} \rangle$ the described scenario means that:
\begin{enumerate}
\item both $\vert \Psi^{(1)} \rangle = \vert HV \rangle$ and $\vert \Psi^{(2)} \rangle = \vert VH \rangle$ states are equally probable (occur with $50\%$ probability),
\item $\vert HV \rangle$ state means that the $\vert H \rangle$-polarized photon propagates in the first channel containing the scattering sample, while $\vert V \rangle$-polarized photon propagates through the second, reference channel (air environment, no scattering medium involved). The $\vert VH \rangle$ state is interpreted in a similar manner.
\end{enumerate}

In this paper, we aim at predicting the evolution of such two-photon polarization-entangled state upon propagation of one of the partner photons through the scattering medium. For this, we express the selected input Bell state in terms of the creation operator $\hat{a}^{\dagger}$~\cite{Wolf1995}:

\begin{equation}
\label{eq:inputBell}
\vert\Psi^+\rangle = \dfrac{1}{\sqrt{2}}\left(\vert HV\rangle+\vert VH\rangle\right) = \dfrac{1}{\sqrt{2}}\left( \hat{a}^{\dagger}_H \vert 0 \rangle \vert V \rangle + \hat{a}^{\dagger}_V \vert 0 \rangle \vert H \rangle \right).
\end{equation}
Here, $\hat{a}^{\dagger}_H \vert 0 \rangle = \vert H \rangle$ and $\hat{a}^{\dagger}_V \vert 0 \rangle = \vert V \rangle$ denote the creation of horizontally and vertically polarized photons from the vacuum state $\vert 0 \rangle$, respectively. We do not use the creation operator to expand polarization states in the second channel because in this work we assume that their probability amplitudes do not change due to the absence of the scattering medium. 

We note that it is possible to define a similar state within Jones vector formalism allowing to bridge both approaches:
\begin{equation}
\label{eq:inputBellInMC}
\mathcal{E} = \dfrac{1}{\sqrt{2}}\left(\boldsymbol\varepsilon_H \otimes \boldsymbol\varepsilon_V + \boldsymbol\varepsilon_V \otimes \boldsymbol\varepsilon_H\right) \rightarrow \vert \Psi^+\rangle .
\end{equation} 
As in Eq.~(\ref{eq:ketDefinition}), the arrow denotes correspondence between field components and respective probability amplitudes which ultimately obeys Eq.~(\ref{eq:densityVScoherencyRelation}). To implement Eq.~(\ref{eq:inputBellInMC}) in the MC model, we launch a photon packet into each channel (see Fig.~\ref{fig:optSetup}) and supplement both launched packets with $\boldsymbol\varepsilon_H$ and $\boldsymbol\varepsilon_V$ polarization states simultaneously, in agreement with Sec.~\ref{sec:polTracing}. Both states at launch are indistinguishable in terms of probability and can be independently traced along each photon packet trajectory with Eq.~(\ref{eqn:SvMC_BSprocedure_expanded}).

In case of scattering, one of the partner photons acquires non-basis polarizaton state vector, which is reflected in the emergence of the non-zero probability amplitudes $\alpha_H, \alpha_V, \beta_H, \beta_V$:

\begin{equation}
\label{eq:scatteredQuantumState}
\begin{aligned}
& \vert\Psi^S\rangle = \dfrac{1}{\sqrt{2}}\left( \alpha_H\hat{a}^{\dagger}_H \vert 0 \rangle \vert V \rangle + \alpha_V\hat{a}^{\dagger}_V \vert 0 \rangle \vert V \rangle + \beta_H\hat{a}^{\dagger}_H \vert 0 \rangle \vert H \rangle + \beta_V\hat{a}^{\dagger}_V \vert 0 \rangle \vert H \rangle \right), \\
& \mathrm{with } \: \vert \alpha_H \vert ^2 + \vert \alpha_V \vert ^2 = 1, \: \vert \beta_H \vert ^2 + \vert \beta_V \vert ^2 = 1.
\end{aligned}
\end{equation}
Within MC, it is possible to follow the evolution of probability amplitudes in terms of Jones vector components. With respect to Eq.~(\ref{eq:X}), each $j$-th photon packet trajectory passing through the turbid medium would result in a unique final polarization state $\mathbf{X}_j$ or $\mathbf{Y}_j$: 
\begin{equation}
\label{eq:XYj}
\mathbf{X}_j = \mathfrak{m}_j\boldsymbol\varepsilon_H + \mathfrak{n}_j\boldsymbol\varepsilon_V, \quad \mathbf{Y}_j = \mathfrak{p}_j\boldsymbol\varepsilon_H + \mathfrak{q}_j\boldsymbol\varepsilon_V.
\end{equation}
Both polarization states are simultaneously evaluated for one photon-packet trajectory, which makes them inherently indistinguishable in agreement with the input Bell state. Expressions (\ref{eq:XYj}) are exclusively valid for the photon packet which passes through the optical channel containing the sample. A second photon packet in the $j$-th pair which is passing through the reference channel is assumed to retain its initial polarization state: either $\boldsymbol\varepsilon_H$ or $\boldsymbol\varepsilon_V$. It is then possible to express the final state of the detected polarization-entangled photon packet pair similarly to Eqs.~(\ref{eq:inputBellInMC}) and (\ref{eq:scatteredQuantumState}): 
\begin{equation}
\label{eq:finalEntangledState}
\begin{aligned}
\mathbf{\mathcal{E}}_j & \propto \mathbf{X}_j \otimes \boldsymbol\varepsilon_V + \mathbf{Y}_j \otimes \boldsymbol\varepsilon_H = \mathfrak{m}_j\boldsymbol\varepsilon_H\otimes \boldsymbol\varepsilon_V + \mathfrak{n}_j\boldsymbol\varepsilon_V \otimes \boldsymbol\varepsilon_V + \mathfrak{p}_j\boldsymbol\varepsilon_H \otimes \boldsymbol\varepsilon_H + \mathfrak{q}_j\boldsymbol\varepsilon_V \otimes \boldsymbol\varepsilon_H = \\ 
& = \left(\begin{array}{c}
    0\\ 
    \mathfrak{m}_j\\ 
    0\\
    \mathfrak{n}_j\\
    \end{array}\right) + \left(\begin{array}{c}
    \mathfrak{p}_j\\ 
    0\\ 
    \mathfrak{q}_j\\
    0\\
    \end{array}\right) = \left(\begin{array}{c}
    \mathfrak{p}_j\\ 
    \mathfrak{m}_j\\ 
    \mathfrak{q}_j\\
    \mathfrak{n}_j\\
    \end{array}\right) \rightarrow \left(\begin{array}{c}
    {\beta_H}_j\\ 
    {\alpha_H}_j\\ 
    {\beta_V}_j\\
    {\alpha_V}_j\\
    \end{array}\right) \propto \vert\Psi^S_j\rangle .
\end{aligned}
\end{equation} 
We explicitly point out that instead of a pure quantum state $\vert \Psi^S_j \rangle$ its Jones-constructed counterpart for the pair of photon packets $\mathcal{E}_j$ is used within MC, which is also reflected in the usage of the arrow denoting surjective relation between $\mathcal{E}_j$ and $\vert \Psi^S_j \rangle$. When there is no scattering medium in the sample channel, all $j=[1 ... N_{ph}]$ coefficients become $\mathfrak{m}_j=1, \mathfrak{n}_j=0, \mathfrak{p}_j=0, \mathfrak{q}_j=1$ and expression (\ref{eq:finalEntangledState}) reduces to the Bell state $\vert \Psi^{+} \rangle$ accurately up to a multiplier. By obtaining coefficients $\mathfrak{m}_j, \mathfrak{n}_j, \mathfrak{p}_j, \mathfrak{q}_j$ from the MC simulation via Eqs.~(\ref{eqn:SvMC_BSprocedure_expanded})--(\ref{eq:X}) we can therefore trace the evolution of the entangled state. Here, we heavily rely on the following facts: (a) each photon packet has a pure state after each scattering event (see Sec.~\ref{sec:polTracing} and Sec.~\ref{sec:bridging}), (b) pure states $\vert \Psi^{(1)} \rangle$ and $\vert \Psi^{(2)} \rangle$, and, correspondingly, $\mathbf{X}_j \otimes \boldsymbol\varepsilon_V$ and $\mathbf{Y}_j \otimes \boldsymbol\varepsilon_H$ are separable (see Eq.~(\ref{eq:qubitState})) and (c) at launch they have $50$\% occurrence probability (see expression for $\vert \Psi^{+} \rangle$).

Finally, we use expressions (\ref{eq:densityMatrixDefinition}) and (\ref{eq:wolfmatrix}) to construct an equivalent of Wolf's coherency matrix for the pure state of the photon packet pair:
\begin{equation}
\label{eq:densityMatrixScattering}
\hat{\rho}_j = \vert\mathcal{E}_j\rangle \langle\mathcal{E}_j\vert \rightarrow \vert \Psi^S_j \rangle \langle \Psi^S_j \vert.
\end{equation}
To be interpreted as a density matrix, $\hat{\rho}_j$ has to be divided by the trace $\mathrm{tr}(\hat{\rho}_j)$ according to Eq.~(\ref{eq:densityVScoherencyRelation}).

\subsection{Matrix averaging over the ensemble of entangled photon packet pairs}
\label{sec:averagingDensityMatrix}

So far we have introduced the coherency and density matrices of a single photon packet pair, i.e. the state of the photon packet pair when a single MC trajectory for the photon packet in the sample channel is considered. This corresponds to finding the final polarization state of single photon packet, which always remains pure, as discussed in Sec.~\ref{sec:polTracing}. This is generally not the case when an ensemble of either photon packets or photon packet pairs is considered. For this reason, we gather statistics for a significant amount of the detected photon packet pairs and address the question of proper averaging over the obtained ensemble to estimate the final state.
 
As mentioned in Sec.~\ref{sec:bridging}, in polarimetric biophotonic applications, where individual photon packets with either $\mathbf{ X}_j$ or $\mathbf{Y}_j$ detected state are considered, one would commonly compute the Stokes vector or, equivalently, $6$ intensity values for each photon packet for horizontal ($H$), vertical ($V$), +45$^{\circ}$ or diagonal ($D$), -45$^{\circ}$ or anti-diagonal ($A$), right- ($R$) and left- ($L$) circular polarization states
\begin{equation}
\begin{array}{cccccc}
    I_H(\mathbf{X}_j), & I_V(\mathbf{X}_j), & I_D(\mathbf{X}_j), & I_A(\mathbf{X}_j), & I_R(\mathbf{X}_j), & I_L(\mathbf{X}_j),\\
    I_H(\mathbf{Y}_j), & I_V(\mathbf{Y}_j), & I_D(\mathbf{Y}_j), & I_A(\mathbf{Y}_j), & I_R(\mathbf{Y}_j), & I_L(\mathbf{Y}_j),\\
    \end{array}
 \label{eq:photonIntensitiesScattered}
\end{equation}
and then average them over the whole ensemble of the detected photon packets~\cite{Lopushenko2024}: 
\begin{equation}
\label{eq:intensityAveraging}
I_{\upphi}=\dfrac{1}{N_{ph}}\sum_{j=1}^{N_{ph}}I_\upphi(\mathbf{X}_j).
\end{equation}
Here, $\upphi$ is the chosen projection state, as in Sec.~\ref{sec:bridging}. This is equivalent to the computation of the ensemble-averaged Wolf's coherency matrix, because the coherency matrix bijectively corresponds to the set of light intensity measurements~\cite{Wolf1959}:
\begin{equation}
\label{eqn:WolfMatrixAveraged}
    \left(\begin{array}{c c}
    \langle E_x E_x^*\rangle & \langle E_x E_y^*\rangle \\ 
    \langle E_y E_x^*\rangle & \langle E_y E_y^*\rangle \\ 
    \end{array}\right) = \dfrac{1}{2}\left(\begin{array}{c c}
    S_0 + S_1 & S_2 + \mathfrak{j}S_3\\ 
    S_2-\mathfrak{j}S_3 & S_0-S_1 \\ 
    \end{array}\right).
\end{equation} 
Here, $\mathfrak{j}$ is the imaginary unit, $S_0=I_H+I_V, S_1=I_H-I_V, S_2=I_D-I_A, S_3=I_R-I_L$ and brackets $\langle \cdot \rangle$ correspond to the field averaging procedure~\cite{Born2019,Wolf1995}. Ensemble averaging (\ref{eq:intensityAveraging}) corresponds to the incoherent superposition of photon packet intensities which may result in a partially polarized light state. By partially polarized state of light we mean that the following relation holds between elements of Stokes vector $\mathbf{S}$: $S_0^2 > S_1^2+S_2^2+S_3^2$. Physically, partially polarized light originates from the field superposition of many light sources with respect to the detector spectral and spatial resolution and finite integration time. To proceed, we note that results established in the paper by James et al.~\cite{Kwiat2001} were obtained for the Stokes vector $\mathbf{S}$ regardless of the fully polarized or partially polarized light state. This ensures that the coherency matrix of the partially polarized light~(\ref{eqn:WolfMatrixAveraged}) and the density matrix of the mixed quantum state~\cite{Kwiat2001,Landau} are related via a matrix trace, as in Eq.~(\ref{eq:densityVScoherencyRelation}).

In the case of two-photon packets, more intensity projections instead of $6$ as defined by~(\ref{eq:photonIntensitiesScattered}) are to be evaluated in order to fully describe the state of the pair. In the experiment, by a proper selection of the independent $\Phi$ state set, i.e. basic orthogonal states to project to, the final density matrix $\hat{\rho}_{\mathrm{}}$ of the quantum state can be reconstructed \cite{Kwiat2001,Altepeter:2004,Alfano2023}. In the studies that we implemented to test our model, we employed quantum state tomography (QST) with one detector per channel and 16 polarization projection combinations (more details are given in S2.2 in Supplementary Material).

While direct reproduction of QST can be implemented in the model, it appears to be redundant. Instead, we demonstrate that it is possible to straightforwardly average an equivalent of Wolf's coherency matrix $\hat{\rho}_j$ defined according to Eq.~(\ref{eq:densityMatrixScattering}) over the ensemble of photon packet pairs thus obtaining a mixed state as a general result. For this purpose, both statistical weight and power of the Rayleigh factor $\Gamma_R$ (see Supplementary Material, S1.2) of the photon packet which passed through the scattering sample have to be accounted for along with its polarization state~\cite{Kuzmin}. Then, by using the concept of Eq.~(\ref{eq:intensityProjection}) we derive how the intensity projection of the photon packet pair onto any allowable photon pair state $\Phi$ is to be evaluated:
\begin{equation}
I_{\Phi}(\hat{\rho}_j) = W_j \langle \Phi \vert \hat{\rho}_j \vert \Phi \rangle \Gamma^{N_j}_R = \langle \Phi \vert W_j \hat{\rho}_j \Gamma^{N_j}_R \vert \Phi \rangle.
\label{eq:intensityProjectionOfThePhotonPair}
\end{equation}
Here, $W_j$ is the detected statistical weight of the $j$-th photon packet which has propagated through the turbid sample, $N_j$ corresponds to the amount of scattering events along the $j$-the photon packet trajectory prior to the detection event, and $\Gamma_R$ is the Rayleigh factor derived from the optical theorem in Born approximation~\cite{TuchinBookMeglinski2013,RoyalSoc2005}. In the second part of the expression, we have used the associativity property for the product between a vector and the real scalar $W_j \Gamma^{N_j}_R$. 

For the ensemble of photon packet pairs, observable intensity projection on the chosen state $\Phi$ is then obtained with the following averaging procedure:
\begin{equation}
I_{\Phi} \propto \sum_{j=1}^{N_{ph}} I_{\Phi}(\hat{\rho}_j) = \sum_{j=1}^{N_{ph}} \langle \Phi \vert W_j\hat{\rho}_j \Gamma^{N_j}_R \vert \Phi \rangle= \langle \Phi \vert \left(\sum_{j=1}^{N_{ph}} W_j\hat{\rho}_j \Gamma^{N_j}_R \right) \vert \Phi \rangle = \langle \Phi \vert \hat{\rho}_\mathrm{avg} \vert \Phi \rangle.
\label{eq:intensityFromDensityMatrix2}
\end{equation}
Here, we have directly applied the summation over the ensemble of the detected photon packet pairs $j=[1 ... N_{ph}]$ to the $W_j \hat{\rho}_j \Gamma^{N_j}_R$ term due to the distributivity of the matrix product with respect to the matrix addition, allowing to introduce $\hat{\rho}_\mathrm{avg}$: a counterpart of Wolf's coherency matrix for an ensemble of photon packet pairs. As opposed to $\hat{\rho}_j$, this averaged matrix in general corresponds to the mixed state, similarly to how the coherency matrix (\ref{eqn:WolfMatrixAveraged}) corresponds to the partially polarized state of light. With account for the relation~(\ref{eq:densityVScoherencyRelation}), we obtain the final expression for the simulated density matrix of the two-photon state which is mixed in the general case:
\begin{equation}
\label{eq:averagedDensityMatrix}
\hat{\rho}_\mathrm{} = \hat{\rho}_\mathrm{avg} / \mathrm{tr}\left( \hat{\rho}_\mathrm{avg} \right).
\end{equation}
Such a matrix comprehensively models the target final two-photon state and thus allows for analysis of the state evolution due to scattering within the turbid medium. Also, by substituting $\upphi$ instead of $\Phi$ and coherency matrix in form of, e.g., $\vert \mathbf{X}_j\rangle\langle \mathbf{X}_j\vert$ instead of $\hat{\rho}_j$, expressions (\ref{eq:intensityProjectionOfThePhotonPair})--(\ref{eq:intensityFromDensityMatrix2}) immediately provide a way to evaluate the observable intensity values for an ensemble of single photon packets. Relations (\ref{eq:finalEntangledState})--(\ref{eq:densityMatrixScattering}) and (\ref{eq:intensityProjectionOfThePhotonPair})--(\ref{eq:averagedDensityMatrix}) are the key expressions of the generalized MC approach and are applied in the following to model the experimental observations.

\section{Results and Discussion}
\label{sec:results}

\subsection{Effect of the turbid tissue-like scattering medium on the density matrix} \label{sec:sampleEffect}

To study the evolution of the polarization-entangled state due to propagation through a scattering medium experimentally and to test the validity of the introduced MC model for state prediction, we implemented the scenario visualized in Fig.~\ref{fig:optSetup}. As samples of scattering medium with different optical properties we selected in-house manufactured tissue-mimicking phantoms with ZnO nanoparticles acting as scattering centers. The scattering properties of the phantoms have been chosen so that the effective thickness $d/l^*$ of the samples gradually reaches $1.0$: $\mu_s^\prime$ = 0.45, 0.96, 1.55, 2.44, and 3.34~mm$^{-1}$. Here, $d$ is the actual thickness of the phantom which approximately equals~$300~\mu$m, $l^*$ is the transport mean free path, and $\mu_s^\prime$ is the reduced scattering coefficient~\cite{Tuchin2015book}. As a reference sample we used a sample fabricated out of the same host material but without scattering centers and thus characterized by negligible scattering coefficient. The detailed description of the samples, realized optical arrangement, measurement protocol, and data processing procedure is given in Sec.~S2 of the Supplementary Material. Below, we summarize the results from our experiments and simulations. 
\begin{figure}[!t]
    \centering    \includegraphics[width=.85\linewidth]{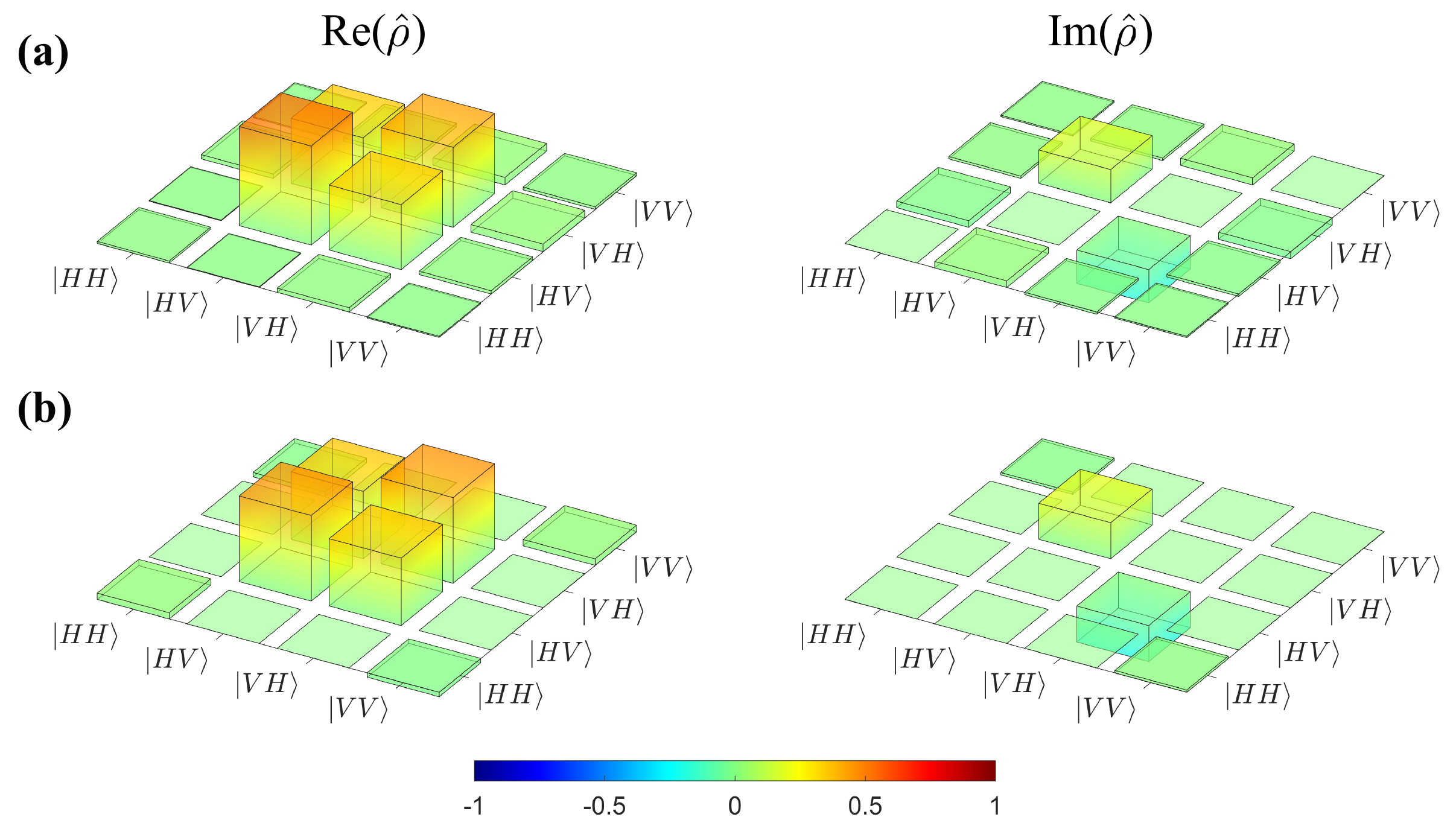}
      \caption{Density matrix of the two-photon state after interaction of one of its partner photons with ZnO-based tissue phantom of $d/l^* \approx$ 1 in one of the arms. (a) Measured and (b) computed with Eq. (\ref{eq:averagedDensityMatrix}) with account for the initial state impurity (see Supplementary Material for details). Simulation parameters are selected to be identical to the measured sample's properties. Theoretical estimate also includes a fit for phase delay equal to $\delta=-\lambda/14$ induced by the possible birefringence of the sample. The obtained fidelity between the measured and simulated matrices is 91\%.}
      \label{fig:scattering}
\end{figure}

\begin{figure}[!b]
    \centering    \includegraphics[width=\linewidth]{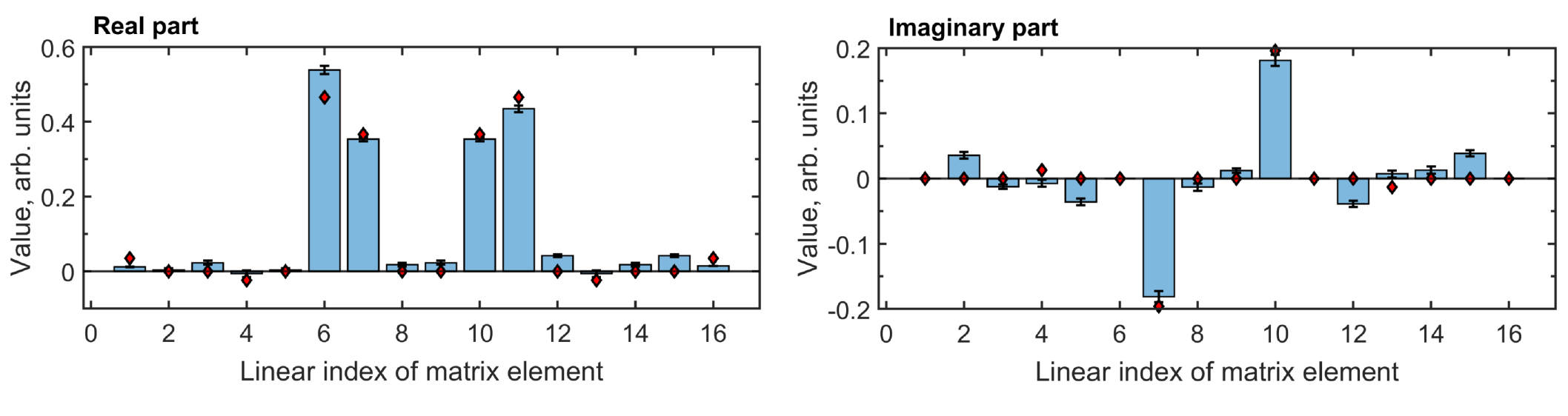}
      \caption{Measured (barplot) and modelled (diamonds) density matrix from Fig.~\ref{fig:scattering} reshaped to vectors. Error bars represent the error estimation of the experimentally retrieved density matrix elements \cite{Kwiat2001}.
      }
      \label{fig:barPlot}
\end{figure}

Figure~\ref{fig:scattering} provides a representative measurement (subfigure a) and simulation (subfigure b) outcome for the sample with the highest reduced scattering coefficient experimentally measured within our study ($\mu_s^\prime$ = 3.34 mm$^{-1}$). This sample is characterized by approximately unit effective thickness $d/l^*$, so that its optical properties are close to those exhibited by real biological tissues, e.g. human epidermis and dermis \cite{Bashkatov:2011}. As can be seen, the matrices qualitatively agree well with each other. For both experiment and simulation outcomes, the coherencies between the $\vert HV \rangle$ and $\vert VH \rangle$ basis states (anti-diagonal elements) transfer to the imaginary part. The corresponding redistribution is estimated to reflect the phase delay of $\lambda/14$ which arises between the vertically and horizontally polarized photons after passing through the tissue phantom. In Supplementary Material we discuss in detail how such interpretation is obtained and the sample-related impact is separated from the phase delays that might be induced by other optics in the path of the photons. The discussed change in the density matrix can indicate the dephasing of the state due to the scattering sample. This effect is more pronounced in the simulated output state where also corner elements attain values slightly exceeding the noise level. They attract the probabilities from the core elements which suggests the appearance of multiple superposition states with different phase relations. For quantitative comparison and better visibility of the presented data we provide the same experimental and simulated outcome as reshaped into a vector and plotted in the same axes in Fig.~\ref{fig:barPlot}. Additionally, we supply the experimental data with the error estimation for retrieval of each density matrix element following the error analysis suggested by James et al. \cite{Kwiat2001}. 

Another minimal discrepancy observed between the measured and simulated matrices is related to the relative amplitudes for the populations of the basis states. The imbalance between the core diagonal elements noticeable in the measurement outcome did not reproduce in the MC simulation. This can, though, be attributed to the probable residual differences between the real probing state and its fitted representation used in simulations. Nevertheless, as we show next, this has no significant impact on the metrics of the quantum state which could be potentially used as diagnostics criteria or monitoring parameters. The values of these metrics for both the measured and calculated state lie in close vicinity as for the provided example as well as for other measured tissue phantoms.

\subsection{Dependence of the polarization-entangled state evolution on the scattering properties of the propagation medium} \label{sec:evolution}

\begin{figure}[!t]
    \centering
    \includegraphics[width=\linewidth]{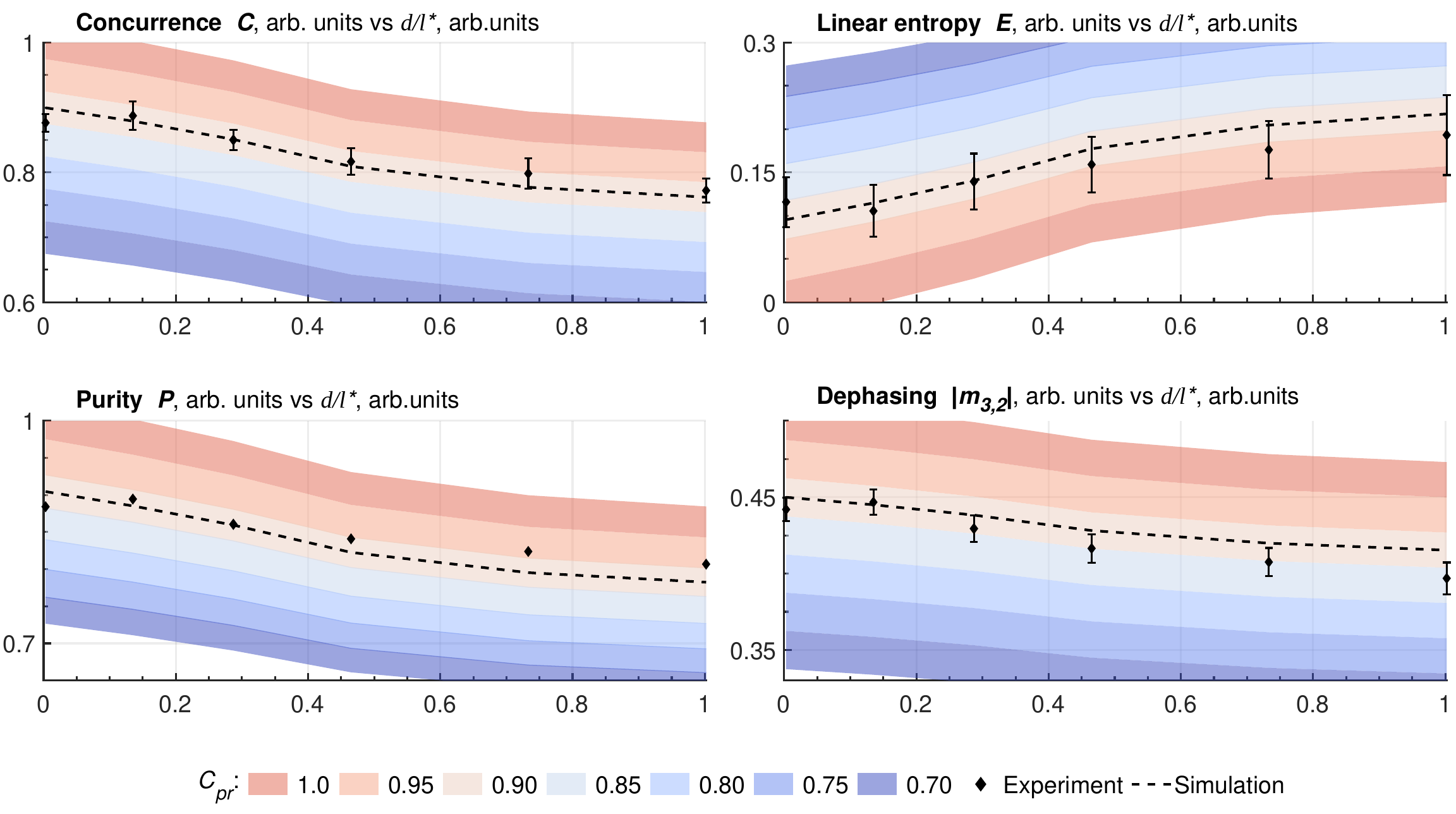}
      \caption{Evolution of the polarization-entangled two-photon state due to interaction with a scattering medium in terms of concurrence, linear entropy, purity, and dephasing of the output state vs effective thickness of the scattering medium~$d/l^*$. Simulation results are shown with shaded stripes for different quality levels of the initial probing state ($C_{pr}$ = 1.0, 0.95, 0.90, 0.85, 0.80, 0.75 and 0.70). Experimentally measured points (diamonds, $C_{pr}$~=~0.88$\pm$0.01) with error estimation \cite{Kwiat2001} and best fitting simulated outcome (dashed line, $C_{pr}$~=~0.90) for $d/l^*$ = 0.003, 0.135, 0.287, 0.465, 0.733, and 1.002.}
    \label{fig:expVSsimOD}
\end{figure}

In order to systematically explore the evolution of the probing polarization-entangled photon pair dependent on the scattering properties of the sample (propagation medium), we monitor several characteristics of the quantum state: concurrence $C$, linear entropy $E$, purity $P$, and dephasing expressed as the magnitude of the anti-diagonal component $|\hat{\rho}_{3,2}|$. Our findings are summarized in Fig.~\ref{fig:expVSsimOD} while fidelities between the measured and simulated density matrices for all studied samples have been obtained in the range from 91\% to 98\%.

Here, the shaded stripes provide an overview of the MC prediction of the output state dependent on the scattering properties of the sample and on the initial quality (entanglement level) of the probing state. These dependencies are shown versus effective thickness of the sample $d/l^*$. The evolution of the probing states of concurrence $C_{pr} =$ 1.00, 0.95, 0.90, 0.85, 0.80, 0.75, and 0.70 is demonstrated. These probing states were generated via Eq.~(\ref{eq:densityMatrixDecompositionIntoBell}) with the corresponding probability weight factors $p_1$ and $p_2$ equal to: $[1, 0]$, $[39/40, 1/40]$, $[19/20, 1/20]$, $[37/40, 3/40]$, $[9/10, 1/10]$, $[7/8, 1/8]$ and $[17/20, 3/20]$ (refer to Supplementary Material for details). Here, the probability factors $p_3$ and $p_4$ are assumed to be equal to zero.

In the experiments, the probing state has been prepared with concurrence (0.88$\pm$0.01). The metrics of the reconstructed density matrices for all measured samples obtained experimentally are represented on corresponding plots with filled diamonds and supplemented with error estimation as per James et al. \cite{Kwiat2001}. The dashed line highlights the MC simulated outcome for the probing state with the concurrence value of 0.90, which was found to be fitting best to the experimental data. Considering the inevitable presence of experimental error and minimal discrepancy between the really generated probing state and its simulated counterpart, the results agree well with each other. 
For all studied metrics, which reflect both the overall properties of the state (concurrence, entropy, purity) as well as direct monitoring of one of the core elements of the density matrix (dephasing), the modelling and experiment correlate with high accuracy. The found behavior correlates also with previously reported theoretical studies~\cite{Velsen:2004} while comprehensive comparison remains the question of future studies, since this would require additional samples of higher scattering to be investigated.

\subsection{Discussion}

The presented results reveal a clear trend of the state evolution for slow loss of entanglement when one of the partner photons passes through a scattering medium. Moreover, we show that this holds valid for different levels of entanglement of the initial state incident on the sample, both predicted by the MC simulations and observed experimentally. The non-ambiguous dependence of the entangled state on the scattering properties of the sample and robustness of the trend to the initial quality of the probing state prove its diagnostic potential and showcases the monitoring of the state evolution as a robust metric for potential quantitative characterization of the medium/sample to be inspected or detected.

The minimal discrepancy between the simulated and experimentally reconstructed states in terms of imbalance in the core diagonal elements of the density matrix described in the previous section can be further improved by more precise matching of the simulated input state to the actually generated state in the experiment. This, though, will require a more complex model of the utilized experimental arrangement and we will address this point in our future studies. Nevertheless, the presented findings explicitly demonstrate the applicability of the entangled states for studying turbid media as well as expands the potential of MC modelling to non-classical states.

The parameters selected for the test set of samples experimentally measured with this study allow to assess the introduced approach of experimental diagnostics and simulation support via MC modelling for a wide range of applications. On one hand, we demonstrated the appropriateness of the method to study samples with scattering coefficient at the range characteristic for real biological tissues (e.g. human epidermis and dermis \cite{Bashkatov:2011}). In agreement to earlier reported works \cite{Shi:2016} the entanglement is preserved on a high level even for samples with the highest reduced scattering coefficient from the test set. This makes the findings of this study particularly relevant for low-flux remote photonics for biomedical diagnostics, and especially the Bell states for probing the birefringence of biomedical samples at low levels of the effect present and/or with potential of the enhanced sensitivity \cite{Pedram:2023, Pedram:2024,Zhang:2024:SciAdv}. On the other hand, the samples with relatively low scattering coefficient match the properties of different conditions of the atmosphere, including the air polluted with particulate matter or featuring presence of water aerosols \cite{Davis2005}. This, in turn, highlights the significance of the proposed method also for such applications as precise remote environmental monitoring, optical communication link maintenance, and reliable quantum optical data transmission. 

The particular impact of the presented study for further development of quantum technologies lies in the inherent scalability of the introduced modelling approach to the simulation of multi-photon problems. In simulation one can introduce the same scattering medium in the second channel of the discussed experimental scenario in a relatively straightforward way. With the cost of the increased computational efforts, one can obtain statistically significant amount of possible trajectories of the photons in both channels and for both $\vert H \rangle$ and $\vert V \rangle$ states. Applying the analytical description of our model the prediction of the output state in this case would similarly arrive at finding the coefficients $\mathfrak{m}, \mathfrak{n}, \mathfrak{p},\mathfrak{q}$. It would be necessary to account here also for coherent interactions between the substates of the decomposition \cite{Safadi:2023}. We will proceed with studying this case in our future research, also experimentally. In addition, the introduced model can be expanded to multi-photon polarization-entangled states. Benefiting from the BSE framework allowing for tracking different polarization states for each single trajectory of the MC photon, we have shown the capabilities of the model on example of a two-photon Bell state in the $\vert H \rangle$ and $\vert V \rangle$ basis. It is, though, possible to track also other types of initial polarization states as well as one can consider several spatial channels of photon propagation and not necessarily only two. We intend to carry out research in this direction in our future work as well.

To sum up, this work studies the scattering of polarization-entangled photons and provides several impacts to the research community. First, we introduced a MC model allowing for prediction and interpretation of the polarization-entangled two-photon state evolution due to scattering in a turbid medium. Second, with the performed experimental studies, we validated the model but also for the first time systematically investigated in practice the effect of the scattering properties of the medium of propagation on the polarization-entangled state and proved its reliability for quantitative and robust monitoring/diagnostics purposes. Next, the selected properties of the scattering samples under study addressed a broad range of practical applications of entangled photons including biomedical diagnostics, environmental monitoring and optical communication. In addition, we discussed the perspective scalability of the introduced approach to multipartite problems. 

With the above mentioned tunability and scalability of the introduced model and the deepened understanding of the polarization-entangled photon interaction with the medium for interpretation and prediction of experimental observation, the presented results are expected to foster the development of quantum-enhanced technologies. This work bridges the fields of optics of scattering media and quantum polarization-based sensing and will accelerate the development of application-oriented quantum technologies.

\section{Summary and Conclusions}
In the present study, we investigated both theoretically and experimentally the scattering of polarization-entangled two-photon states when one of the partner photon experiences scattering in a turbid medium of known optical properties. We have generalized the framework of the Monte Carlo polarization tracing approach based on the iterative solution to the Bethe-Salpeter equation to the case of two-photon state description, and for the first time have systematically expressed this approach in terms of both coherency and density matrices. We experimentally validated the model and systematically studied a series of scattering samples. Their optical properties in terms of effective thickness range from the values relevant for particulate matter polluted atmosphere to those of biological tissues. We have demonstrated the reliability of the polarization-entangled photons for quantitative and robust monitoring/diagnostics purposes. The proposed approach is expected to foster the establishment of optimal measurement algorithms using non-classical states of light, exploration of the limitations of quantum polarimetry, and form the basis for elaboration of numerous photonic applications leveraging the quantum technologies. We also discussed the possibility to expand the introduced modelling approach to two-photon scattering as well as to simulation of multi-photon polarization-entangled states.

\section*{Funding}
This article is based upon work from COST Action CA21159 - Understanding interaction light - biological surfaces: possibility for new electronic materials and devices (PhoBioS), supported by COST (European Cooperation in Science and Technology). The work has also been partly funded by German Ministry of Education and Research (project "QUANCER", FKZ 13N16441), Academy of Finland (grant project 325097) and UK Department for Science Innovation and Technology in partnership with the British Council. V.B. thanks for funding of this work also to ProChance-career program of the Friedrich Schiller University Jena.

\section*{Acknowledgments}

We would like to thank Prof. Fabian Steinlechner and Mr. Purujit Singh Chauhan for providing the source of polarization-entangled photon pairs and Ms. Luosha Zhang for her help in alignment of the measurement system. 

\section*{Disclosures}
The authors declare no conflicts of interest.

\section*{Data Availability Statement}
Data supporting the findings of this study are available from the corresponding authors upon reasonable request.

\section*{Supplementary Material}
See Supplementary Material document for supporting content. 

%%%%% References %%%%%
\bibliography{refs}   % bibliography data in refs.bib
\bibliographystyle{spiejour}   % makes bibtex use spiejour.bst

%%%%% Biographies of authors %%%%%

\vspace{2ex}\noindent\textbf{Vira R. Besaga, Dr.-Ing.} studied at Chernivtsi National University (Ukraine) and then at Ruhr University Bochum (Germany), where she earned her doctorate degree in 2020. In 2021 she joined the Friedrich Schiller University Jena (Germany) as a research fellow in quantum imaging and sensing. Her research focuses on classical and quantum-enhanced optical metrology and sensing, covering areas such as quantitative phase imaging, holographic microscopy, and polarimetry for applications in biomedical diagnostics, technical inspection, and fundamental research. 

\vspace{2ex}\noindent\textbf{Ivan V. Lopushenko, Dr.} received his doctoral degree in physics and mathematics from Lomonosov Moscow State University (Russia) for the development of semi-analytic numerical models in electromagnetic scattering and nonlocal nanoplasmonics in 2019. Since 2020 he joined the University of Oulu (Finland) as a PostDoc in light scattering simulations. His research interests are rooted in mathematical physics and involve different aspects of scattering-related problems, from the fundamental level to the implementation of relevant high-performance numerical tools for biophotonics, mesoscopic electrodynamics and quantum state tomography. 

\vspace{2ex}\noindent\textbf{Oleksii Sieryi} Biography of the author is not available.

\vspace{2ex}\noindent\textbf{Alexander Bykov, D.Sc. (Tech.)} is an adjunct professor (Docent) in biophotonics and sensors technologies at the Optoelectronics and Measurement Techniques Unit (OPEM), University of Oulu, Finland. In 2010, he received his doctoral degree. Since 2019, he has been the leader of the Biophotonics Group at OPEM. His scientific interests are in the areas of non-invasive optical diagnostics of biotissues, the theory of light propagation in scattering media, and numerical simulation of light transport. He is a senior member of OPTICA.

\vspace{2ex}\noindent\textbf{Frank Setzpfandt, Dr. rer. nat.} is a research group leader for quantum optics at the Friedrich Schiller University in Jena, Germany. He studied physics at the Friedrich Schiller University, from which he also received his PhD in 2012. Before returning to Jena, he spent time as a PostDoc at the Australian National University in Canberra. His research is concentrating on nonlinear frequency conversion and the generation of photon pairs, quantum imaging and sensing, and integrated quantum optics. 

\vspace{2ex}\noindent\textbf{Igor Meglinski, Prof. Dr.} is a professor in quantum biophotonics and biomedical engineering at the College of Engineering and Physical Sciences at Aston University (UK). His research interests lie at the interface between physics, optics, and biomedical engineering, focusing on the development of new non-invasive imaging/diagnostic techniques and their application in medicine and biology, pharmacy, environmental monitoring, food sciences and health care industries. He is a chartered physicist (CPhys) and chartered engineer (CEng), senior member of IEEE, Fellow of Institute of Physics, FRMS, Fellow SPIE, and Fellow OPTICA (formerly Optical Society of America).

%\end{spacing}

\end{document}